\documentclass[aps,prd,twocolumn,showpacs,nofootinbib]{revtex4-1}

\usepackage{graphicx}
\usepackage[colorlinks=true, linkcolor=blue, citecolor=blue, urlcolor=blue]{hyperref}

\begin{document}

\title{Impacts of the intrinsic charm content of the proton on the $\Xi_{cc}$ hadroproduction at a fixed target experiment at the LHC}

\author{Gu Chen$^1$}
\email{email:speecgu@gzhu.edu.cn}
\author{Xing-Gang Wu$^2$}
\email{email:wuxg@cqu.edu.cn}
\author{Shuai Xu$^2$}
\email{shuaixu@cqu.edu.cn}

\address{$^1$School of Physics and Electronic Engineering, Guangzhou University, Guangzhou 510006, People's Republic of China\\
$^2$Department of Physics, Chongqing University, Chongqing 401331, People's Republic of China}

\date{\today}

\begin{abstract}

In the present paper, we present detailed discussions on the hadronic production of $\Xi_{cc}$ at a fixed target experiment at the LHC (After@LHC). The charm quarks in hadron could be either extrinsic or intrinsic. By using the BHPS model as the intrinsic charm distribution function in proton, we observe that even if by setting the proportion of finding the intrinsic charm in a proton as $A_{\rm in}=1\%$, total cross sections for the $g+c$ and $c+c$ production mechanisms shall be enhanced by nearly two times. Thus the number of $\Xi_{cc}$ events to be generated at the After@LHC can be greatly enhanced. Since the total cross sections and differential distributions for the $\Xi_{cc}$ production at the After@LHC are sensitive to the value of $A_{\rm in}$, the After@LHC could be a good platform for testing the idea of intrinsic charm.  \\


\end{abstract}

\maketitle

\section{Introduction}

Stimulating by the observation of the doubly charmed baryon $\Xi_{cc}^{++}$ by the LHCb collaboration~\cite{Aaij:2017ueg}, people have shown many new interests on the doubly heavy baryons. More measurements are assumed to be done at the LHCb Upgrade II~\cite{Bediaga:2018lhg}. In the past decades, in addition to its decay properties, many theoretical works have been done for the production of the doubly heavy baryons at various high-energy colliders~\cite{Falk:1993gb, Kiselev:1994pu, Baranov:1995rc, Berezhnoy:1998aa, Gunter:2001qy, Braguta:2002qu, Braaten:2003vy, Ma:2003zk,Li:2007vy, Yang:2007ep, Zhang:2011hi, Jiang:2012jt, Jiang:2013ej, Martynenko:2013eoa, Chen:2014frw, Yang:2014tca, Yang:2014ita, Martynenko:2014ola, Leibovich:2014jda, Leibovich:2014jda, Brown:2014ena, Zheng:2015ixa, Trunin:2016uks, Berezhnoy:2016wix, Brodsky:2017ntu, Huan-Yu:2017emk, Yao:2018zze, Niu:2018ycb}.

There are three important mechanisms for the production of $\Xi_{cc}$ at the high-energy hadronic colliders such as LHC and Tevatron, which are through the gluon-gluon fusion ($g+g$), the gluon-charm collision ($g+c$), and the charm-charm collision ($c+c$), respectively. Those production mechanisms are pQCD calculable, since the intermediate gluon should be hard enough to generate a hard $c\bar{c}$ pair in the final state. For the $(g+c)$ and $(c+c)$ production mechanisms, one usually treats the incident charm quarks as ``extrinsic" ones, which are perturbatively generated by gluon splitting according to the DGLAP evolution~\cite{Gribov:1972ri, Altarelli:1977zs, Dokshitzer:1977sg}. The hadronic production of $\Xi_{cc}$ with ``extrinsic" charm mechanism has been discussed in Refs.~\cite{Chang:2006eu, Chen:2014hqa, Chen:2018koh}. Those works show that the $(g+c)$ mechanism dominates over the conventionally consider $(g+g)$ fusion mechanism in small $p_t$ region~\footnote{In large $p_t$ region, the cross section shall be highly suppressed by the charm quark distribution function; This explains why the gluon-gluon mechanism alone is usually adopted for analyzing the measurements with large $p_t$ cut.}, and thus it is important for the fixed-target experiments such as the SELEX experiment at the Tevatron and the suggested fixed target experiment at the LHC (After@LHC) due to the measured $\Xi_{cc}$ $p_t$ could be very small~\cite{Brodsky:2012vg, Hadjidakis:2018ifr, Lansberg:2012wj, Lansberg:2012sq, Lansberg:2013wpx}.

\begin{figure}[htb]
\includegraphics[width=0.4\textwidth]{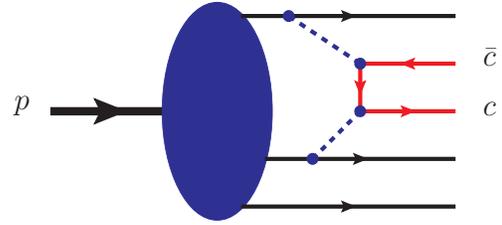}
\caption{Typical Feynman diagrams for the intrinsic mechanism through nonperturbative fluctuations of the proton state to five-quark Fock state. The dashed lines stand for soft interactions. }
\label{intr}
\end{figure}

In addition to the ``extrinsic" ones, the incident $c$-quarks may also be ``intrinsic" ones, which are correlated to the non-perturbative fluctuations of nucleon state to the five-quark state, as shown in Fig.~\ref{intr}. This idea has been proposed firstly by Brodsky {\it et al.}, and the BHPS model has been raised for estimating the intrinsic $c$-quark distribution in nucleon~\cite{Brodsky:1980pb, Brodsky:1981se, Brodsky:2015fna}. Lately, many more phenomenological studies have been done to illustrate the non-perturbative charm in nucleon, e.g., the meson-baryon model~\cite{Navarra:1995rq, Hobbs:2013bia}, the sea-like model~\cite{Pumplin:2005yf}, and etc.. Because the proportion of the intrinsic charm components in nucleon is small, which is only up to $\sim 1\%$, the intrinsic charm usually gives negligible contribution in most of the high-energy processes. At present, due to lack of experimental measurements, definite conclusion on the existence of intrinsic charm is still missing.

It has been found that the $\Xi_{cc}$ events generated at the SELEX are much more sensitive to the intrinsic charm than those at the hadronic colliders as LHC and the Tevatron~\cite{Chang:2006xp, Koshkarev:2016rci, Koshkarev:2016acq, Groote:2017szb}. There is hope to confirm the intrinsic components in proton by measuring the events in specific kinematic regions, such as small $p_t$ region. The SELEX experiment has already been shut down and its puzzle on $\Xi_{cc}$ observation, e.g., its measured production rate is much larger than most of the theoretical predictions~\cite{Mattson:2002vu, Ocherashvili:2004hi}, remains unresolved. The intrinsic charm production mechanism may solve this puzzle~\cite{Brodsky:2017ntu}. And we still need more accurate fixed-target experimental data to clarify the issue. At the LHC, when the incident proton beam energy rises up to 7 TeV, the proposed After@LHC will run with a center-of-mass energy around 115 GeV. With a much higher luminosity and higher collision energy, the After@LHC will become a much better fixed-target experiment for studying the properties of the doubly heavy baryons. It is thus interesting to investigate how and to what degree the intrinsic charm affects the $\Xi_{cc}$ production at the After@LHC.

The remaining parts of the paper are organized as follows. In Sec.II, we present the calculation technology for the hadronic production of $\Xi_{cc}$. In Sec.III, we present our numerical results and discussions for various $\Xi_{cc}$ hadroproduction mechanisms, and show how the intrinsic charm affects the cross sections. Sec.IV is reserved for a summary.

\section{Calculation Technology}

Within the perturbative QCD factorization formula, the total cross section for the hadronic production of $\Xi_{cc}$ can be factorized as follows,
\begin{widetext}
\begin{eqnarray}
\sigma(H_1+H_2 \to \Xi_{cc}+X)&=& \int dx_1 dx_2 \bigg\{f^{g}_{H_{1}}(x_{1},\mu) f^{g}_{H_{2}}(x_{2},\mu) \otimes\hat{\sigma}_{g+g\rightarrow \Xi_{cc}}(x_{1},x_{2},\mu)\nonumber\\
&+& \sum_{i,j=1,2;i\neq j} f^{g}_{H_{i}} (x_{1},\mu)\left[f^{c}_{H_{j}}(x_{2},\mu) - f^{c}_{H_{j}}(x_{2},\mu)_{\rm SUB} \right] \otimes \hat{\sigma}_{g+c\rightarrow \Xi_{cc}}(x_{1},x_{2},\mu)\nonumber\\
&+& \sum_{i,j=1,2;i\neq j} f^{c}_{H_{i}}(x_{1},\mu)  f^{c}_{H_{j}}(x_{2},\mu) \otimes \hat{\sigma}_{cc\rightarrow \Xi_{c+c}}(x_{1},x_{2},\mu)+ \cdots \bigg\}, \label{facform}
\end{eqnarray}
\end{widetext}
where we have implicitly set the factorization scale and renormalization scale to be the same, $\mu_F=\mu_R=\mu$. $f_H^a$ ($a=(g,c)$) is parton distribution function (PDF) of the corresponding parton $a$ in the incident hadron $H$. $f^{c}_{H}(x,\mu)_{\rm SUB}$ is the subtraction term to avoid double counting problem between the $(g+g)$ and $(g+c)$ production mechanisms~\cite{Aivazis:1993kh, Aivazis:1993pi, Olness:1997yc, Amundson:2000vg}, which is defined as,
\begin{eqnarray}\label{subtraction}
f^{c}_{H}(x,\mu)_{\rm SUB}& \equiv & f^{g}_{H}(x,\mu) \otimes f^{c}_g(x,\mu) \nonumber\\
&=& \int^1_{x}\frac{dy}{y}f^{c}_g(y,\mu) f^{g}_{H}\left(\frac{x}{y},\mu\right)
\end{eqnarray}
with
\begin{eqnarray} \label{subtraction2}
f^c_g(x,\mu) &=& \frac{\alpha_s(\mu)}{2\pi} \ln\frac{\mu^2}{m^2_c}P_{g\to q}(x) \nonumber\\
&=&\frac{\alpha_s(\mu)} {2\pi}\ln\frac{\mu^2}{m^2_c} \cdot \frac{1}{2}(1-2x+2x^2).
\end{eqnarray}

By taking the intrinsic charm component into account, the PDF $f_H^a$ can be expressed as,
\begin{eqnarray}\label{apsg}
f_H^a(x,\mu)&=& f_H^{a,0}(x,\mu)+f_H^{a,\rm in}(x,\mu),
\end{eqnarray}
where $f_H^{a,0}$ is the PDF without intrinsic charm effect, and $f_H^{a,\rm in}(x,\mu)$ is the new term introduced by the intrinsic charm effect.

The PDF at any other scale can be obtained by applying the DGLAP equations with the known PDF $f_H^{a,\rm in}(x,2m_c)$ at the initial scale $2 m_c$, i.e.,~\cite{Field:1989uq}
\begin{widetext}
\begin{eqnarray}
f^{c,\rm in}_H(x,\mu)&=& \int_x^1 \frac{dy}{y}  \left\{f^{c,\rm in}_H(x/y,2m_c) \frac{[-\ln(y)]^{a_c\kappa-1}}{\Gamma(a_c \kappa)}\right\}+ \nonumber\\
&& \kappa\int_x^1\frac{dy}{y} \int_y^1 \frac{dz}{z} \left \{f^{c,\rm in}_H(y/z,2m_c) \frac{[-\ln(z)]^{a_c\kappa-1}}{\Gamma(a_c\kappa)} P_{\Delta c}(x/y) \right \}+{\cal O}(\kappa^2),
\label{intrc}
\end{eqnarray}
\begin{eqnarray}
f^{g,\rm in}_H(x,\mu)&=& \frac{2\kappa}{a_g-a_c}\int_x^1 \frac{dy}{y}\int_{a_c}^{a_g}da\int_y^1 \frac{dz}{z}\left\{f^{c,\rm in}_H(z,2m_c) \frac{[-\ln(z)]^{a\kappa-1}}{\Gamma(a\kappa)}P_{c\to gc}(x/y)\right\}+{\cal O }(\kappa^2),
\label{intrg}
\end{eqnarray}
\end{widetext}
with
\begin{eqnarray}\label{intrg1}
&&a_g = 6,  \ a_{c}=\frac{8}{3}, \beta_0=11-2n_f/3, \nonumber\\
&&\kappa=\frac{2}{\beta_0} \ln \left(\frac{\alpha_s(2m_c)}{\alpha_s(\mu)}\right), \nonumber\\
P_{\Delta c}(x) &=& \frac{4}{3} \left[\frac{1+x^2}{1-x}+\frac{2}{\ln x}+\left(\frac{3}{2}-2\gamma_E\right)\delta(1-x)\right], \nonumber\\
&& P_{c\to gc} = \frac{4}{3}\left[\frac{1+(1-x)^2}{x}\right].
\end{eqnarray}
In doing the numerical analysis, we adopt the BHPS model~\cite{Brodsky:1980pb} for the PDF $f^{c,\rm in}_H(x,2m_c)$ as a typical one to discuss the intrinsic charm's effect, e.g.,
\begin{eqnarray}\label{bhps}
&&f^{c,\rm in}_H(x,2m_c)   \nonumber\\
&&= 6  x^2 \xi \left[6x(1+x) \ln x + (1-x) (1 + 10x + x^2) \right] \; ,
\end{eqnarray}
where the parameter $\xi$ is fixed by the probability of finding the intrinsic charm quark, which satisfies the normalization condition as,
\begin{displaymath}
A_{\rm in}\equiv \int_0^1 f^{c, \rm in}_H(x,2m_c)\; dx=\xi \times 1\% \;.
\end{displaymath}
The probability for finding intrinsic $c/\bar{c}$-component in proton at the fixed low-energy scale $2m_c$ is assumed to be less than $1\%$~\cite{Brodsky:1980pb, Brodsky:1981se}, and we set a broader range of $\xi\in[0.1,1]$ to do the discussion.

Many effects have been paid to the intrinsic charm (IC) PDF~\cite{Pumplin:2007wg, Nadolsky:2008zw, Martin:2009iq, Dulat:2013hea, Jimenez-Delgado:2014zga, Lyonnet:2015dca, Ball:2016neh, Hou:2017khm}, which are usually fixed via global fitting of experimental data. For example, the CTEQ group, firstly suggested the CTEQ6.5C PDF version~\cite{Pumplin:2007wg} by carrying out a series of global fits with varying magnitudes of IC components. That is, the intrinsic charm component is characterized by the first moment of the $c$-quark and $\bar{c}$-antiquark momentum distributions,
\begin{equation}
\langle x\rangle_{c+\bar{c}} = \int_0^1 x [c(x)+\bar{c}(x)]dx,   \label{icteq}
\end{equation}
where the distributions $c(x)$ and $\bar{c}(x)$ depend on the IC models such as the BHPS model (\ref{bhps}), the Meson-Cloud Model (MCM) with the IC arises from virtual low-mass meson+baryon components, e.g., $\bar{D}^0\Lambda^+_c$, in a proton, and the sea-like model with IC is assumed to behave as the light flavor sea quarks, e.g. $c(x) = \bar{c}(x)$ is proportional to $\bar{d}(x)+\bar{u}(x)$ with an overall charm mass suppression. Lately, the CTEQ group improved it as CTEQ6.6C~\cite{Nadolsky:2008zw} IC PDF version by taking both the BHPS and the sea-like models into account with moderate and large IC contributions as $1\%$ and $3.5\%$ (corresponding to $\langle x\rangle_{c+\bar{c}}=0.57\%$ and $2\%$, respectively), which then improved as CT10C~\cite{Dulat:2013hea} and CT14C~\cite{Hou:2017khm} by taking more data into consideration. As another example, the MSTW group issued the MSTW2008 IC PDF version~\cite{Martin:2009iq} by dealing with the IC component under the general-mass variable flavour number scheme.  And recently the NNPDF group developed a model independent NNPDF3IC IC version~\cite{Ball:2016neh}, whose input parameters are based on a NLO calculation and are fixed via a global fitting of experimental data of deep inelastic structure functions.  \\

\section{Numerical results and discussions}\label{results}

The doubly charmed baryon $\Xi_{cc}$ can be produced by first perturbatively forming a $(cc)$ pair via $g+g\to (cc)+\bar{c}\bar{c}$, $g+c\to (cc)+\bar{c}$ or $c+c\to (cc)+g$ channels, then forming a bound $(cc)$-diquark state either in spin-triplet and color anti-triplet state $(cc)_{\bf\bar{3}}[^3S_1]$ or in spin-singlet and color sextuplet state $(cc)_{\bf 6}[^1S_0]$, and finally, hadronizing into the $\Xi_{cc}$ baryon. To be the same as those of Ref.\cite{Chen:2014hqa}, we take the probability for a $(cc)$-pair to transform into the $\Xi_{cc}$-baryon as $|\Psi_{cc}(0)|^2 =0.039$ GeV$^3$, $M_{\Xi_{cc}}=3.50$ GeV with $m_c=M_{\Xi_{cc}}/2$. We take the CT14LO PDF version~\cite{ct14lo}, which is issued by the CTEQ group, as the input for the PDF $f_H^{a,0}(x,\mu)$ without intrinsic charm effect.

\begin{widetext}
\begin{center}
\begin{table}[htb]
\begin{tabular}{|c|c|c|c|c|c|c|}
\hline
~~~-~~~ & \multicolumn{2}{|c|}{$\sigma_{g+g}$ (pb)} &\multicolumn{2}{|c|} {$\sigma_{g+c}$(pb)} & \multicolumn{2}{|c|} {$\sigma_{c+c}$ (pb)} \\
\hline
- & ~~~$(cc)_{\bar{{\bf 3}}}[^3S_1] $~~~ & ~~~$(cc)_{{\bf 6}}[^1S_0]$~~~  & ~~~$(cc)_{\bar{{\bf 3}}}[^3S_1] $~~~ & ~~~$(cc)_{{\bf 6}}[^1S_0] $~~~  & ~~~$(cc)_{\bar{{\bf 3}}}[^3S_1] $~~~ & $(cc)_{{\bf 6}}[^1S_0] $~~~  \\
\hline
$A_{\rm in}=0$~& $7.44\times 10^{2}$ & $1.35\times 10^{2}$ & $3.07\times 10^{3}$ & $3.34\times 10^{2}$ & 1.02 & $4.12\times 10^{-2}$    \\
\hline
$A_{\rm in}=0.1\%$~& $7.47\times 10^{2}$ & $1.35\times 10^{2}$ & $3.31\times 10^3$ & $3.59\times 10^{2}$ & 1.09 & $4.38\times 10^{-2}$ \\
\hline
$A_{\rm in}=0.3\%$~& $7.49\times 10^{2}$ & $1.36\times 10^{2}$ & $3.76\times 10^{3}$ & $4.07\times 10^{2}$ & 1.24 & $4.98\times 10^{-2}$  \\
\hline
$A_{\rm in}=1\%$~& $7.55\times 10^{2}$ & $1.37\times 10^{2}$ & $5.32\times 10^{3}$ & $5.78\times 10^{2}$ & 1.79 & $7.16\times 10^{-2}$  \\
\hline
\end{tabular}
\caption{Total cross sections of the $\Xi_{cc}$ production at the After@LHC with different intrinsic charm component corresponding to different choices of $A_{\rm in}$, which are $0$, $0.1\%$, $0.3\%$, and $1\%$, respectively. $A_{\rm in}=0$ means no intrinsic charm component has been taken into consideration. $p_t>0.2\;\rm GeV$.}
\label{tcro}
\end{table}
\end{center}
\end{widetext}

In the literature, a generator GENXICC~\cite{Chang:2007pp, Chang:2009va, Wang:2012vj} has been programmed, which can be conveniently used for simulating the $\Xi_{cc}$ events at the hadronic colliders. Our numerical calculations shall be done by using the generator GENXICC with proper changes to include both the extrinsic and intrinsic charm effects in the charm and gluon PDFs. The probability of finding the intrinsic charm in proton is set as $A_{\rm in}=0$, $0.1\%$, $0.3\%$, and $1\%$, respectively, where $A_{\rm in}=0$ corresponds to the extrinsic mechanism. We have implicitly taken a small transverse momentum ($p_t$) cut for the $\Xi_{cc}$ events, i.e., $p_t > 0.2$ GeV, which is the same as the SELEX and could also be adopted by the fixed-target experiment After@LHC.

As an overall impression, we present the total cross sections for the $\Xi_{cc}$ production at the After@LHC via the $(g+g)$, $(g+c)$, and $(c+c)$ production mechanisms in Table \ref{tcro}, where the results for $(cc)_{\bf\bar{3}}[^3S_1]$ and $(cc)_{\bf 6}[^1S_0]$ are presented. Table~\ref{tcro} shows that for each production channels, the intermediate $(cc)_{\bf 6}[^1S_0]$ can also give sizable contributions, e.g. its production cross sections for $(g+g)$, $(g+c)$, and $(c+c)$ production mechanisms are about $18\%$, $11\%$ and $4\%$ of the corresponding $(cc)_{\bf\bar{3}}[^3S_1]$ cross sections. Table~\ref{tcro} also shows how the total cross sections vary with the increment of intrinsic charm components in proton, which shall give sizable contributions to the $(g+c)$ and $(c+c)$ mechanisms. For example, even if there is only one-in-one-thousand probability to find the intrinsic charm component in proton, e.g. $A=0.1\%$, the total cross sections for $(g+c)$ and $(c+c)$ mechanisms shall be increased by about $7\%$.

\subsection{$\Xi_{cc}$ production via the $(g+g)$ fusion mechanism}

\begin{table}[htb]
\begin{tabular}{|c|c|c|c|c|}
\hline
- & $p_t \ge 2$ GeV~ & $p_t \ge 4$ GeV & $p_t \ge 6$ GeV & $p_t \ge 8$ GeV \\
\hline
$\sigma_{g+g}^{(cc)_{\bar{{\bf 3}}}[^3S_1]}$ & $2.71\times10^2$ & $3.21\times10^1$ & 3.59 & $4.81\times10^{-1}$ \\
\hline
$\sigma_{g+g}^{(cc)_{{\bf 6}}[^1S_0]}$ & $5.85\times10^{1}$ & 9.06 & 1.21 & $1.80\times10^{-1}$ \\
\hline
\end{tabular}
\caption{Total cross sections (in unit pb) for the $\Xi_{cc}$ production via $(g+g)$ channel at the After@LHC under different $p_t$ cuts, where we have set $A_{\rm{in}}=1\%$.}
\label{ptcgg}
\end{table}

\begin{table}[htb]
\begin{tabular}{|c|c|c|c|}
\hline
- & $|y|<1$ & $|y|<2$ & $|y|<3$ \\
\hline
 $\sigma_{g+g}^{(cc)_{\bar{{\bf 3}}}[^3S_1]}$ & $4.97\times10^2$ & $7.28\times10^2$ & $7.57\times10^{2}$  \\
\hline
$\sigma_{g+g}^{(cc)_{{\bf 6}}[^1S_0]}$ & $8.92\times10^{1}$ & $1.32\times10^{2}$ & $1.37\times10^{2}$ \\
\hline
\end{tabular}
\caption{Total cross sections (in unit pb) for the $\Xi_{cc}$ production via $(g+g)$ channel at the After@LHC under different $y$ cuts, where we have set $A_{\rm{in}}=1\%$ and $p_t>0.2\;\rm GeV$.}
\label{ycgg}
\end{table}

As for $(g+g)$ fusion mechanism, total cross sections with intrinsic charm $A_{\rm in}=1\%$ under various kinematic cuts are presented in Tables~\ref{ptcgg} and~\ref{ycgg}. It's found that the impacts of intrinsic charm on the $(g+g)$ channel is less than $2\%$ even by setting $A_{\rm in}=1\%$. There are nearly $96\%$ $\Xi_{cc}$ events to be generated in small $p_t$ region, $p_t\in[0,4\;\rm GeV]$, and about $66\%$ $\Xi_{cc}$ events for $|y|\le1$. Thus for a fixed-target experiment as After@LHC, in which small $p_t$ events can be detected, a more accurate production information on $\Xi_{cc}$ can be achieved.

\begin{figure}[htb]
\includegraphics[width=0.5\textwidth]{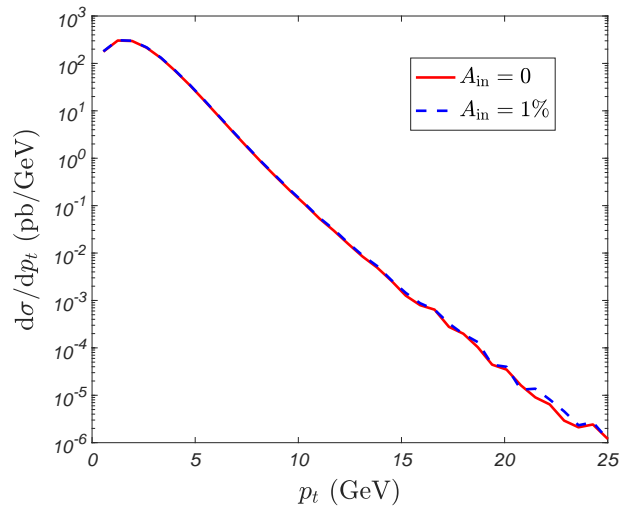}
\caption{Comparison of the $p_t$ distributions for the hadroproduction of $\Xi_{cc}$ with and without intrinsic charm, $A_{\rm in}=1\%$ and $A_{\rm in}=0$, via the $g+g$ production mechanism at the After@LHC. Here contributions from various intermediate diquark states have been summed up. }  \label{cptgg}
\end{figure}

\begin{figure}[htb]
\includegraphics[width=0.5\textwidth]{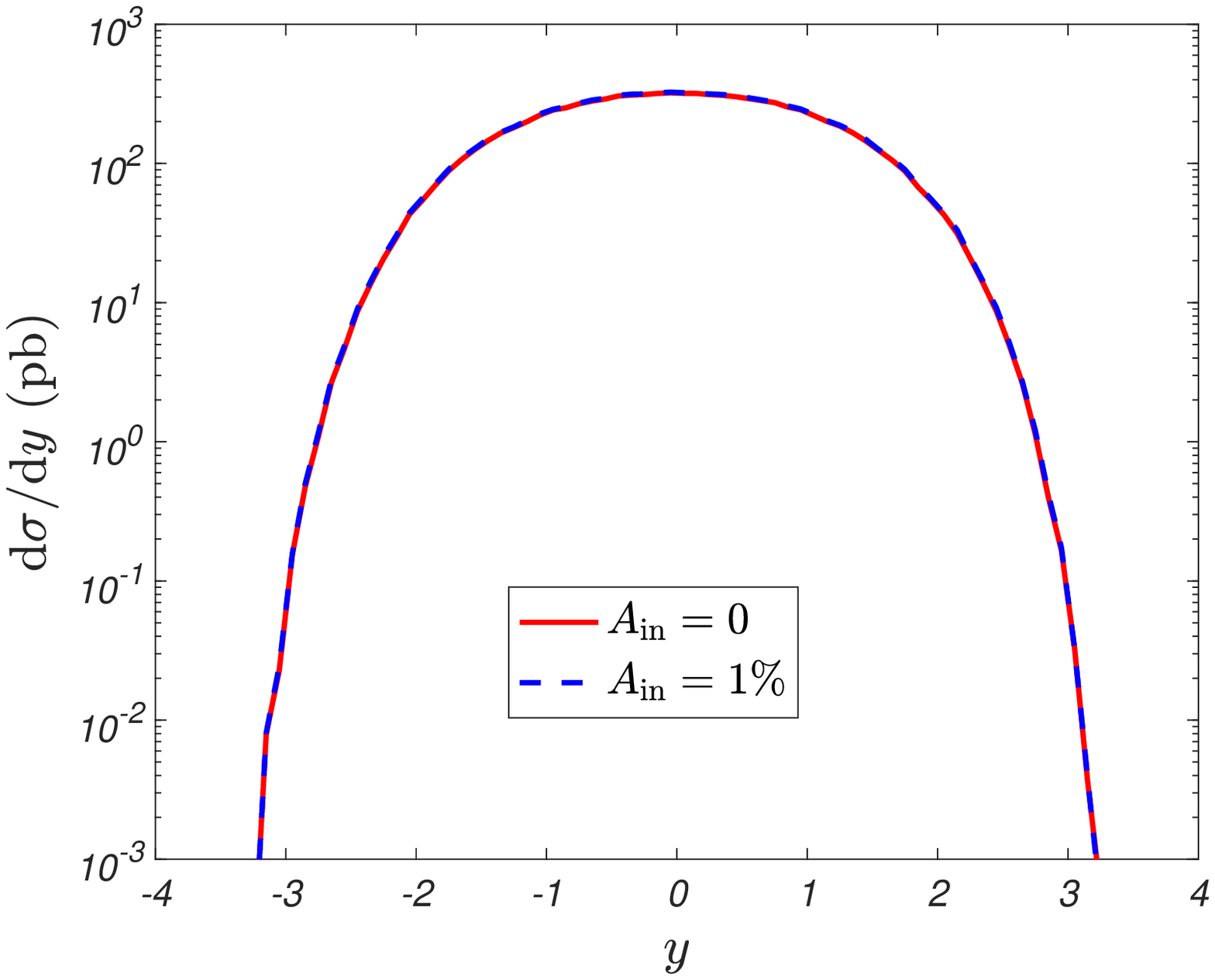}
\caption{Comparison of the $y$ distributions for the hadroproduction of $\Xi_{cc}$ with and without intrinsic charm, $A_{\rm in}=1\%$ and $A_{\rm in}=0$, via the $g+g$ production mechanism at the After@LHC. Here contributions from various intermediate diquark states have been summed up. $p_t>0.2\;\rm GeV$.}  \label{crapgg}
\end{figure}

\begin{figure}[htb]
\includegraphics[width=0.5\textwidth]{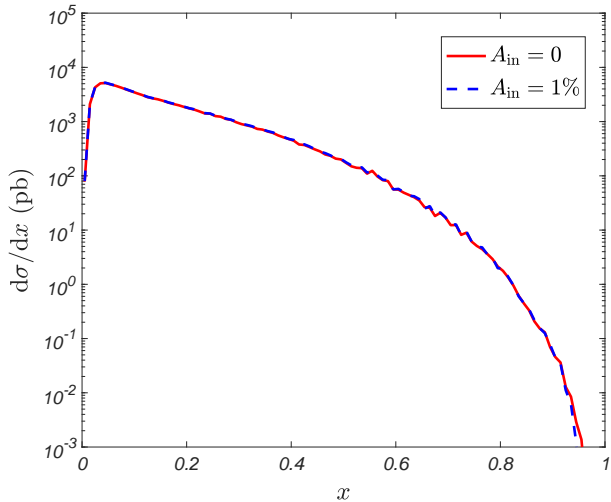}
\caption{Comparison of the $x$ distributions for the hadroproduction of $\Xi_{cc}$ with and without intrinsic charm, $A_{\rm in}=1\%$ and $A_{\rm in}=0$, via the $g+g$ production mechanism at the After@LHC. Here contributions from various intermediate diquark states have been summed up. $p_t>0.2\;\rm GeV$.}  \label{xgg}
\end{figure}

For the differential productions of $\Xi_{cc}$, we investigate the differential distributions with respect to the $p_t$ and $y$ as presented in Figs.~\ref{cptgg} and~\ref{crapgg}, respectively. Both the cases with and without intrinsic charm are plotted, in which the contributions from $(cc)_{\bf\bar{3}}[^3S_1]$ and $(cc)_{\bf 6}[^1S_0]$ diquark states have summed up. In those figures, the solid and the dashed lines stand for the differential distributions without and with intrinsic charm, which correspond to $A_{\rm in}=0$ and $A_{\rm in}=1\%$, respectively. Fig.~\ref{cptgg} shows that the $p_t$-distribution drops quickly with the increment of $p_t$. Fig.\ref{crapgg} shows that there is a small plateau within $|y|\le1.5$ for the $\Xi_{cc}$ production via the $(g+g)$ channel. In Fig.~\ref{xgg}, we plot the $x$ distributions of $\Xi_{cc}$ production with and without intrinsic charm via the $(g+g)$ scheme.

\begin{figure}[htb]
\includegraphics[width=0.5\textwidth]{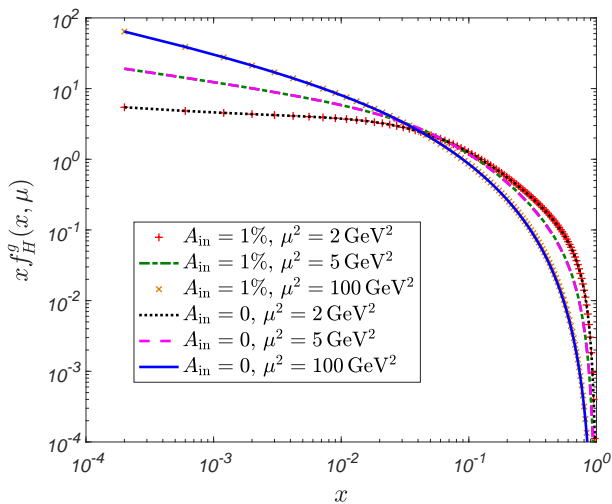}
\caption{The gluon PDF with and without intrinsic charm, $A_{\rm in}=1\%$ and $A_{\rm in}=0$, at different scales ($\mu^2$).}  \label{pdfg}
\end{figure}

Figs.~\ref{cptgg} and~\ref{crapgg} indicate that the $p_t$ and $y$ shapes of $\Xi_{cc}$ change very slightly in whole $p_t$ or $y$ region by taking the intrinsic charm component into consideration. This is due to the fact that the impacts of intrinsic charm to the gluon PDF, as expressed by Eq.~(\ref{intrg}), is small. We present a comparison of the gluon PDF with and without intrinsic charm effects in Fig.~\ref{pdfg}, where three typical scales, $\mu^2=2\;\rm GeV^2$, $5\;\rm GeV^2$, $100\;\rm GeV^2$, are adopted. The nearly coincidence of the two curves with and without intrinsic charm under various scales, indicating the effect of intrinsic charm to the gluon PDF is negligible.

\subsection{$\Xi_{cc}$ production via $(g+c)$ and $(c+c)$ channels with extrinsic charm mechanism}

In addition to the $(g+g)$ channel, the gluon-charm $(g+c)$ and the charm-charm $(c+c)$ interactions are important for a sound prediction of the $\Xi_{cc}$ hadronic production. In this subsection, we study the hadronic production properties of $\Xi_{cc}$ via the $(g+c)$ and $(c+c)$ channel at the After@LHC experiment, where the $c$ quark is extrinsic one only.

To see more explicitly how these channels affect the $\Xi_{cc}$ production cross sections, we define a ratio ${\cal R}$ based on the cross section of the frequently considered channel $g+g \to \Xi_{cc}(cc)_{\bar{{\bf 3}}}[^3S_1]+\bar{c}+\bar{c}$, i.e.,
\begin{eqnarray}
{\cal R} = \frac{\sigma_{\rm tot}}{\sigma_{g+g \to \Xi_{cc}(cc)_{ \bar{{\bf 3}}}[^3S_1]}},
\label{Rr}
\end{eqnarray}
where $\sigma_{\rm tot}$ stands for the total cross sections of all the concerned production mechanisms and intermediate diquark states. The values of ${\cal R}$ shall be shown in Table~\ref{Rval}, where $A_{\rm in}=0$ indicates the extrinsic charm components, whose contribution is large in comparison to the $(g+g)$-mechanism, e.g., ${\cal R}=5.8$ for $A_{\rm in}=0$.

In Table~\ref{tcro}, the results for $A_{\rm in}=0$ are cross sections for extrinsic charm mechanisms. For the $(g+c)$ channel, total cross sections from the diquark state $(cc)_{\bar{{\bf 3}}}[^3S_1]$ are about $9$ times bigger than those from $(cc)_{\bf 6}[^1S_0]$. For the $(c+c)$ channel, total cross sections from the diquark state $(cc)_{\bar{{\bf 3}}}[^3S_1]$ are about $10$ times bigger than those from $(cc)_{\bf 6}[^1S_0]$. By summing up different diquark contributions, the relative importance of the cross sections among different production channels is
\begin{displaymath}
\sigma_{g+g}^{A_{\rm in}=0}: \sigma_{g+c}^{A_{\rm in}=0}: \sigma_{c+c}^{A_{\rm in}=0} \simeq 8.3\times10^{2} : 3.2\times10^{3} : 1.
\end{displaymath}
We observe that the cross section for the $(g+c)$-channel is dominant over that of $(c+c)$-channel by about three orders, which is about four times of the cross section of the $(g+g)$-channel. This confirms the necessity for including the charm-initiated channels in the calculations.

\subsection{The intrinsic charm effects in $\Xi_{cc}$ production via $(g+c)$ and $(c+c)$ channels}

In this subsection we show how the total production cross sections are altered by further taking into account the intrinsic charm.

By varying the intrinsic component $A_{\rm in}$ form $0.1\%$ to $1\%$, the cross sections of $(g+c)$ and $(c+c)$ channels have been presented in Table~\ref{tcro}. The cross sections of $(g+c)$ and $(c+c)$ channels are enhanced by about $7.5\%$ to $75\%$ with increment of the intrinsic charm component $A_{\rm in}\in[0.1\%, 1\%]$. More explicitly, if taking the intrinsic charm component as $A_{\rm in}=1\%$, the relative importance of cross sections among different channels is
\begin{displaymath}
\sigma^{A_{\rm in}=1\%}_{g+g}: \sigma^{A_{\rm in}=1\%}_{g+c}: \sigma^{A_{\rm in}=1\%}_{c+c}\simeq 4.8\times10^{2} : 3.2\times10^{3} : 1.
\end{displaymath}
Comparing with the extrinsic case, we find that the relative importance of $(g+c)$ and $(c+c)$ channels are enhanced by taking the intrinsic charm into consideration.

\begin{table}[htb]
\begin{center}
\begin{tabular}{|c|c|c|c|c|}
\hline
 & ~$A_{\rm in}=0$~ & ~$A_{\rm in}=0.1\%$~ & ~$A_{\rm in}=0.3\%$~ & ~$A_{\rm in}=1\%$~  \\
\hline
 ${\cal R}$ & $5.8$ & $6.1$ & $6.7$ & $9.0$ \\
\hline
\end{tabular}
\caption{The ${\cal R}$ values defined in Eq.~(\ref{Rr}) at the After@LHC with various choices of $A_{\rm in}$. $A_{\rm in}=0$ indicates that only the extrinsic mechanisms are considered. $p_{t}>0.2$~GeV. }
\label{Rval}
\end{center}
\end{table}

We present the ${\cal R}$ ratios under different choices of intrinsic charm components in Table~\ref{Rval}. Table~\ref{Rval} shows the production cross section under extrinsic mechanisms shall be highly affected by the intrinsic charm, e.g., when $A_{\rm in}=1\%$, the ${\cal R}$ ratio shall be increased by $55\%$.

\begin{figure}[htb]
\includegraphics[width=0.5\textwidth]{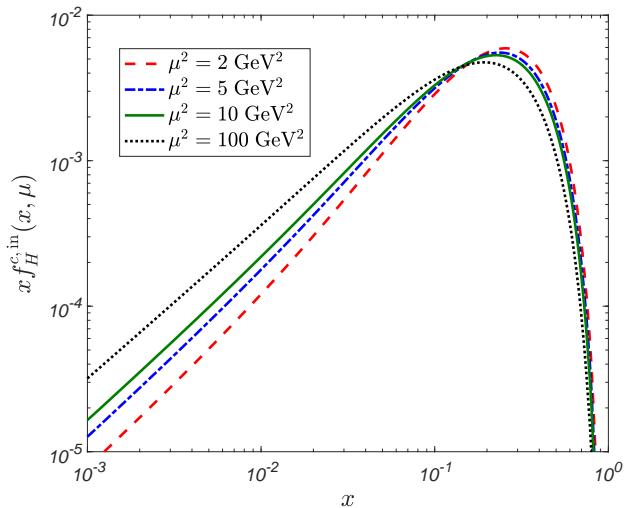}
\caption{Scale evolution of the intrinsic charm PDF defined in Eq.~(\ref{intrc}). $A_{\rm in}=1\%$.}
\label{pdfic}
\end{figure}

\begin{figure}[htb]
\includegraphics[width=0.5\textwidth]{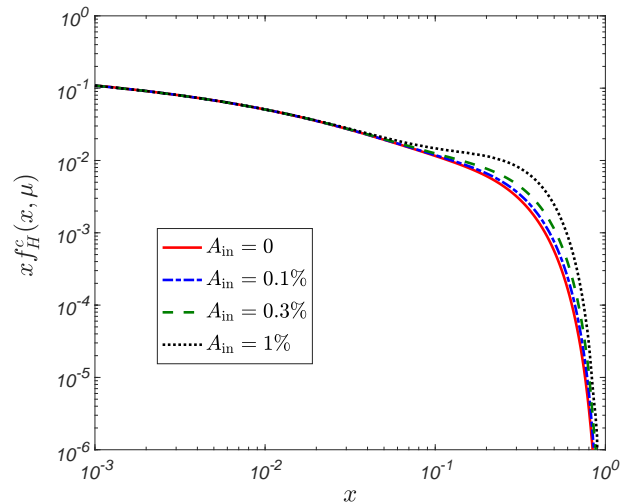}
\caption{Total charm PDF defined in Eq.~(\ref{apsg}) with various intrinsic charm components characterized by $A_{\rm in}=0\sim 1\%$. $\mu^2=5\;\rm GeV^2$.}
\label{pdfc}
\end{figure}

To account for these points, we illustrate how the intrinsic charm component affects the charm PDF. First, we present the $x$-distribution of intrinsic charm with $A_{\rm in}=1\%$ under several typical scales in Fig.~\ref{pdfic}. Fig.~\ref{pdfic} shows the intrinsic charm PDF increases in small $x$ region and decreases in high $x$, whose peak slightly moves with varying scales. Second, we present the total charm PDF, defined in Eq.~(\ref{apsg}), with various intrinsic charm components in Fig.~\ref{pdfc}. It shows the total charm PDF has a small humped behavior around $x\sim 0.3$. This peaked behavior explains the strong enhancement of the intrinsic charm to the $\Xi_{cc}$ production via $(g+c)$ and $(c+c)$ channels at the After@LHC. Thus, the intrinsic charm, if exists in hadrons, shall play an important role in the hadronic production of $\Xi_{cc}$.

Summing up the contributions from different intermediate diquark states and various production channels together, we obtain $\sigma_{\rm tot}^{A_{\rm in}=0}=4.28\times10^{3}$ pb and $\sigma^{A_{\rm in}=1\%}_{\rm tot}=6.79\times10^{3}$ pb. If the integrated luminosity at the After@LHC reaches $0.05\,{\rm fb}^{-1}$ or $2\,{\rm fb}^{-1}$ per operation year~\cite{Brodsky:2012vg}, the $\Xi_{cc}$ events to be generated at the After@LHC shall be about $2.1\times10^5$ or $8.6\times10^6$ per operation year for $A_{\rm in}=0$. If setting $A_{\rm in}=1\%$, the $\Xi_{cc}$ events shall be greatly increased to $3.4\times10^5$ or $1.4\times10^7$ per operation year. Thus to compare with the hadronic production at the LHC which usually adopts a larger $p_t$ cut, the fixed-target experiment After@LHC could provide a better platform for studying the $\Xi_{cc}$ properties and for testing the existence of intrinsic charm.

\begin{table}[htb]
\begin{tabular}{|c|c|c|c|c|}
\hline
 & $p_t \ge 2$ GeV~ & $p_t \ge 4$ GeV & $p_t \ge 6$ GeV & $p_t \ge 8$ GeV \\
\hline
$\sigma_{g+c}^{(cc)_{\bar{{\bf 3}}}[^3S_1]}$ & $1.26\times10^3$ & $8.93\times10^1$ & 8.75 & 1.18 \\
\hline
$\sigma_{g+c}^{(cc)_{{\bf 6}}[^1S_0]}$ & $1.47\times10^{2}$ & $1.52\times10^{1}$ & 1.78 & $2.73\times10^{-1}$ \\
\hline
$\sigma_{g+c}^0$ & $8.04\times10^{2}$ & $5.81\times10^{1}$ & 5.56 & $7.48\times10^{-1}$ \\
\hline
$\sigma_{c+c}^{(cc)_{\bar{{\bf 3}}}[^3S_1]}$ & 1.79 & 1.79 & 1.54 & $3.38\times10^{-1}$  \\
\hline
$\sigma_{c+c}^{(cc)_{{\bf 6}}[^1S_0]}$ & $7.16\times10^{-2}$ & $7.16\times10^{-2}$ & $5.89\times10^{-2}$ & $1.05\times10^{-2}$ \\
\hline
$\sigma_{c+c}^0$ & 1.06 & 1.06 & $8.96\times10^{-1}$ & $1.70\times10^{-1}$  \\
\hline
\end{tabular}
\caption{Total cross sections (in unit pb) for the $\Xi_{cc}$ production at the After@LHC under different $p_t$ cuts, where we have set $A_{\rm{in}}=1\%$. The total cross sections for $A_{\rm in}=0$ are presented as a comparison, e.g., $\sigma^0$ stands for the $\Xi_{cc}$ production without intrinsic charm, where contributions of different diquark configuration have been summed up.}
\label{ptc}
\end{table}

\begin{table}[htb]
\begin{tabular}{|c|c|c|c|}
\hline
 & $|y|<1$ & $|y|<2$ & $|y|<3$ \\
\hline
 $\sigma_{g+c}^{(cc)_{\bar{{\bf 3}}}[^3S_1]}$ & $2.28\times10^3$ & $4.50\times10^3$ & $5.27\times10^{3}$  \\
\hline
$\sigma_{g+c}^{(cc)_{{\bf 6}}[^1S_0]}$ & $2.54\times10^{2}$ & $4.94\times10^{2}$ & $5.78\times10^{2}$ \\
\hline
$\sigma_{gc}^0$ & $1.98\times10^{3}$ & $3.16\times10^{3}$ & $3.39\times10^{3}$  \\
\hline
$\sigma_{c+c}^{(cc)_{\bar{{\bf 3}}}[^3S_1]}$ & 1.43 & 1.79 & 1.79  \\
\hline
$\sigma_{c+c}^{(cc)_{{\bf 6}}[^1S_0]}$ & $5.66\times10^{-2}$ & $7.14\times10^{-2}$ & $7.16\times10^{-2}$ \\
\hline
$\sigma_{cc}^0$ & $8.92\times10^{-1}$ & 1.06 & 1.06  \\
\hline
\end{tabular}
\caption{Total cross sections (in unit pb) for the $\Xi_{cc}$ production at the After@LHC under different $y$ cuts, where we have set $A_{\rm{in}}=1\%$. The total cross sections for $A_{\rm in}=0$ are presented as a comparison, e.g., $\sigma^0$ stands for the $\Xi_{cc}$ production without intrinsic charm, where contributions of different diquark configuration have been summed up. $p_{t}>0.2$ GeV.}
\label{yc}
\end{table}

For convenience of comparing with the future experimental measurements, we present total cross sections under various kinematic cuts in Tables~\ref{ptc} and \ref{yc}, where we have set $A_{\rm in}=1\%$. Tables~\ref{ptc} shows the results for typical transverse momentum cuts, $p_t\ge2\,\rm GeV$, $p_t\ge4\,\rm GeV$, $p_t\ge6\,\rm GeV$, and $p_t\ge8\,\rm GeV$, respectively. There are over $98\%$ contributions are concentrated in small $p_t$ region $[0,\;4\rm GeV]$. Table~\ref{yc} shows the results under three rapidity cuts, $|y|\le1$, $|y|\le2$, and $|y|\le3$.

\begin{table}[htb]
\begin{center}
\begin{tabular}{|c|c|c|c|c|c|}
\hline
 & $p_t \ge 2$ GeV~ & $p_t \ge 4$ GeV & $p_t \ge 6$ GeV & $p_t \ge 8$ GeV \\
\hline
$\varepsilon_{g+c}\,(p_{t\rm cut})$ & 75$\%$ & 80$\%$ & 89$\%$ & 94$\%$ \\
\hline
$\varepsilon_{c+c}\,(p_{t\rm cut})$ & 75$\%$ & 75$\%$ & 78$\%$ & 105$\%$ \\
\hline
\end{tabular}
\caption{The values of $\varepsilon_i(p_{t\rm cut})$ defined in Eq.~(\ref{varep}) for the hadronic production of $\Xi_{cc}$ at the After@LHC with $A_{\rm in}=1\%$.}
\label{eppt}
\end{center}
\end{table}

\begin{table}[htb]
\begin{center}
\begin{tabular}{|c|c|c|c|}
\hline
$y_{\rm cut}$ & $|y|\le1$ & $|y| \le 2$ & $|y| \le 3$ \\
\hline
$\zeta_{g+c}\,(y_{\rm cut})$ & 28$\%$ & 58$\%$ & 73$\%$ \\
\hline
$\zeta_{c+c}\,(y_{\rm cut})$ & 67$\%$ & 76$\%$ & 76$\%$ \\
\hline
\end{tabular}
\caption{The values of $\zeta_{i}(y_{\rm cut})$ defined in Eq.~(\ref{zeta}) for the hadronic production of $\Xi_{cc}$ at the After@LHC with $A_{\rm in}=1\%$. $p_{t}>0.2$ GeV.}
\label{zey}
\end{center}
\end{table}

To see how the kinematic cuts affect the intrinsic charm contributions, we introduce two variables $\varepsilon_i\,(p_{t\rm cut})$ and $\zeta_i\,(y_{\rm cut})$:
\begin{eqnarray}\label{varep}
\varepsilon_i\,(p_{t\rm cut}) = \frac{\sigma_i(p_t\geq p_{t\rm cut})-\sigma^0_i(p_t\geq p_{t\rm cut})}{\sigma^0_i(p_t\geq p_{t\rm cut})}\times 100\%,
\end{eqnarray}
and
\begin{eqnarray}\label{zeta}
\zeta_i\,(y_{\rm cut}) = \frac{\sigma_i(|y| \le y_{\rm cut})-\sigma^0_i(|y| \le y_{\rm cut})}{\sigma^0_i(|y| \le y_{\rm cut})}\times 100\%,
\end{eqnarray}
where $i=g+c$ or $i=c+c$ stands for the contribution from the production channel $g+c\to \Xi_{cc}$ or $c+c\to \Xi_{cc}$, respectively. $\sigma_i^0$ is the cross section without intrinsic charm and $\sigma_i$ denotes that with $A_{\rm in}=1\%$, in which contributions of different diquark configuration have been summed up. The values of $\varepsilon_i$ and $\zeta_i$ with different $p_t$ cuts and $y$ cuts are given in Tables~\ref{eppt} and \ref{zey}. From Table~\ref{eppt}, one can see that the relative importance of the intrinsic charm increases with increment of $p_t$ cuts, e.g., $\varepsilon_{g+c}$ varies from $75\%$ to $94\%$ and $\varepsilon_{c+c}$ varies from $75\%$ to $105\%$ by taking the $p_t$ cut from $2$ GeV to $8$ GeV. As shown in Table~\ref{zey}, the ratio of intrinsic charm contributions $\zeta_i$ significantly increase from $28\%$ to $73\%$ for the $(g+c)$ channel and mildly increase from $67\%$ to $73\%$ for the $(c+c)$ channel with the increment of $y_{\rm cut}$.

\begin{figure}[htb]
\includegraphics[width=0.5\textwidth]{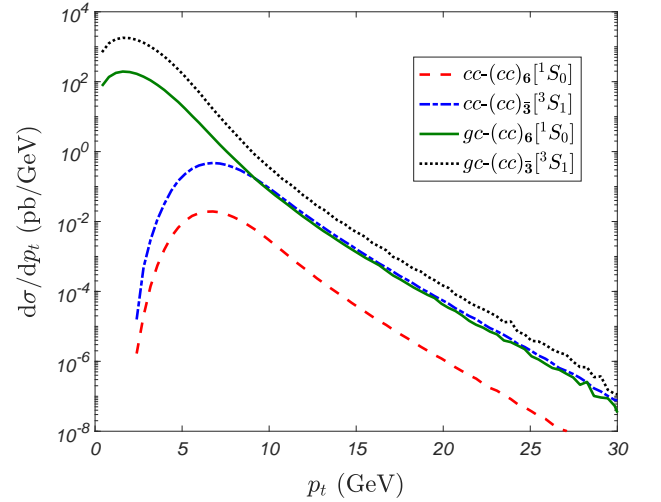}
\caption{The $p_t$ distributions of $\Xi_{cc}$ for various intermediate diquark states at the After@LHC with intrinsic charm component as $A_{\rm in}=1\%$, in which no $y$ cut has been applied.}  \label{pt}
\end{figure}

\begin{figure}[htb]
\includegraphics[width=0.5\textwidth]{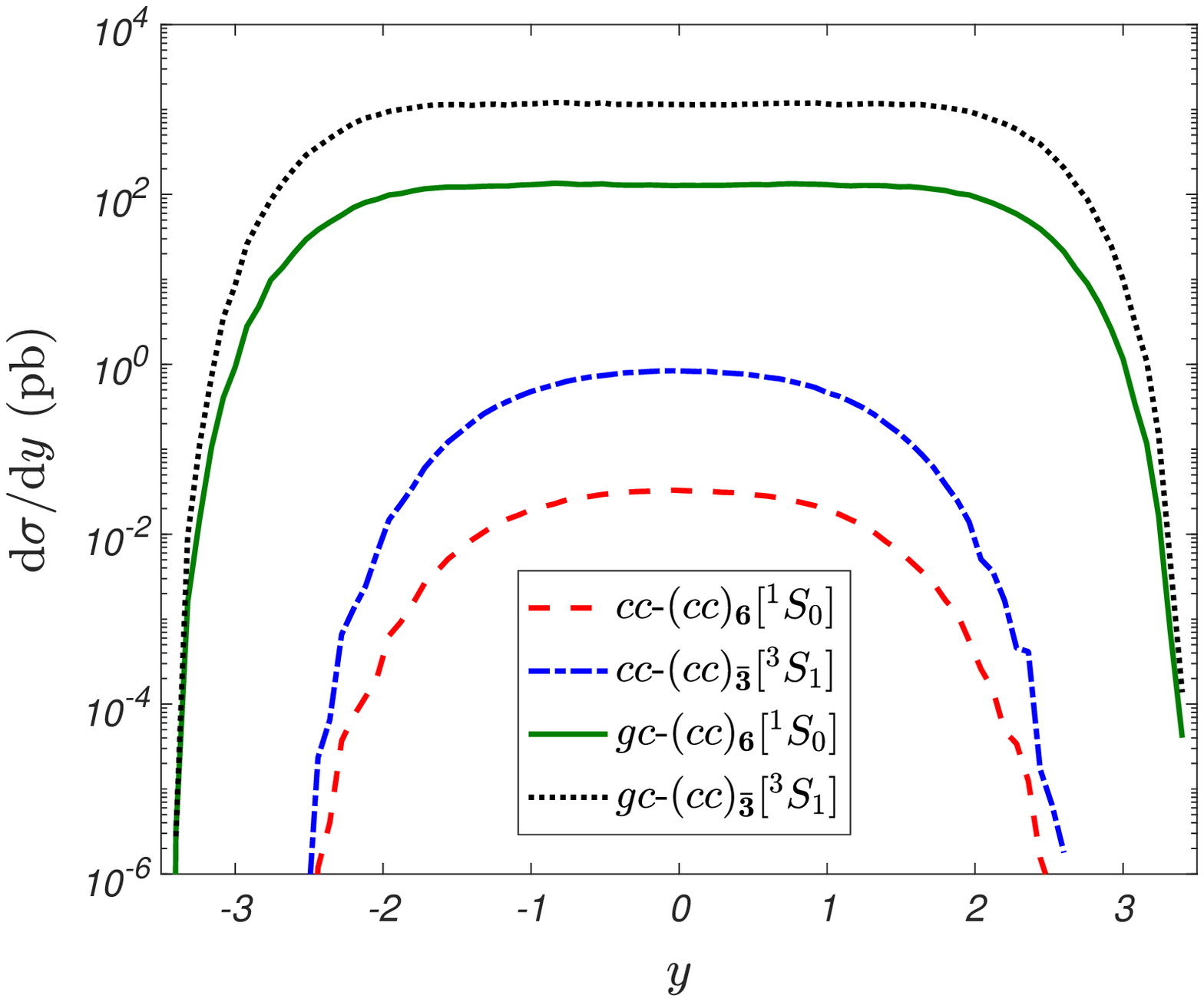}
\caption{The $y$ distributions of $\Xi_{cc}$ for various intermediate diquark states at the After@LHC with intrinsic charm component as $A_{\rm in}=1\%$. $p_t>0.2\;\rm GeV$.}  \label{rap}
\end{figure}

\begin{figure}[htb]
\includegraphics[width=0.5\textwidth]{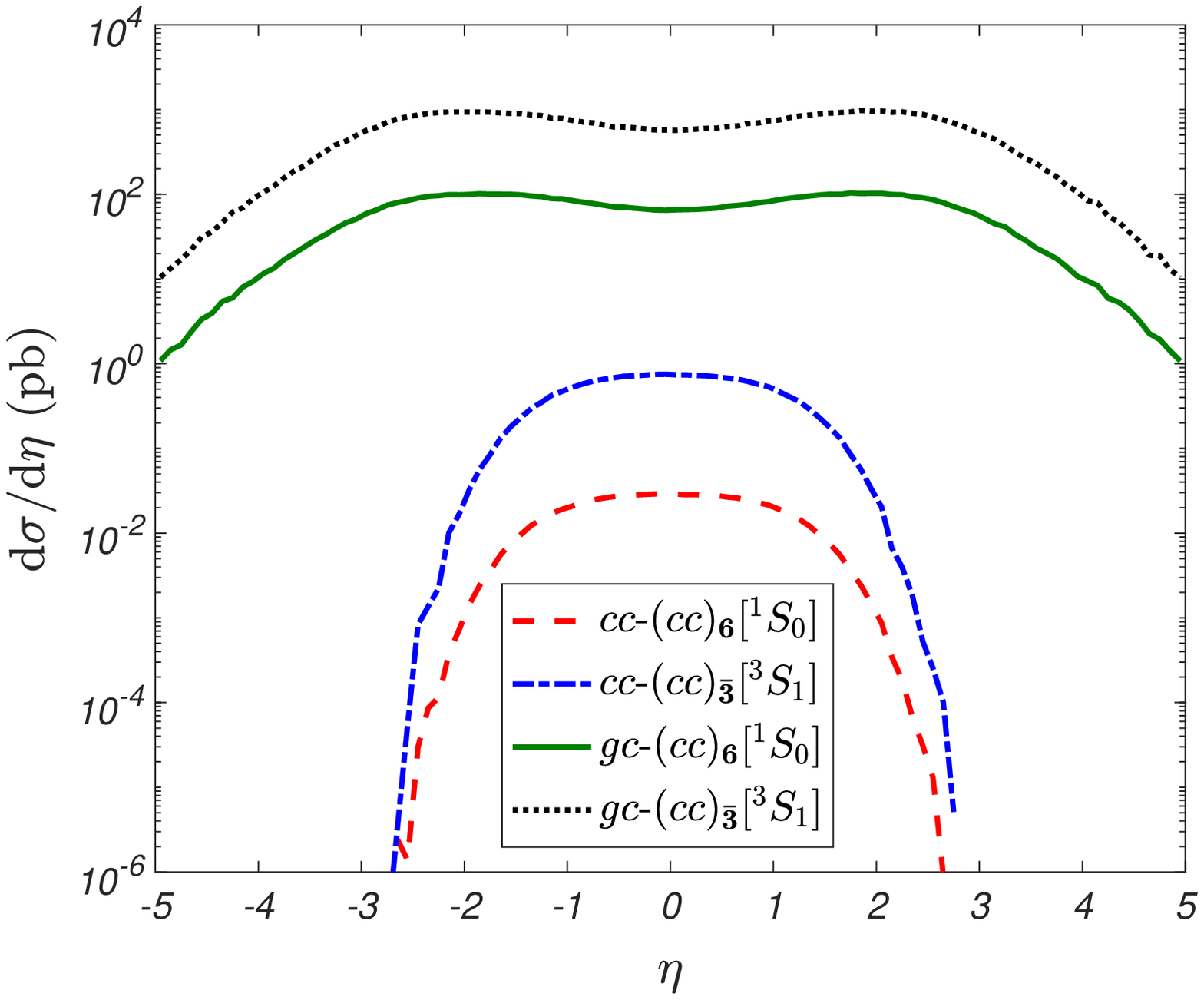}
\caption{The $\eta$ distributions of $\Xi_{cc}$ for various intermediate diquark states at the After@LHC with intrinsic charm component as $A_{\rm in}=1\%$. $p_t>0.2\;\rm GeV$. }  \label{psrap}
\end{figure}

\begin{figure}[htb]
\includegraphics[width=0.5\textwidth]{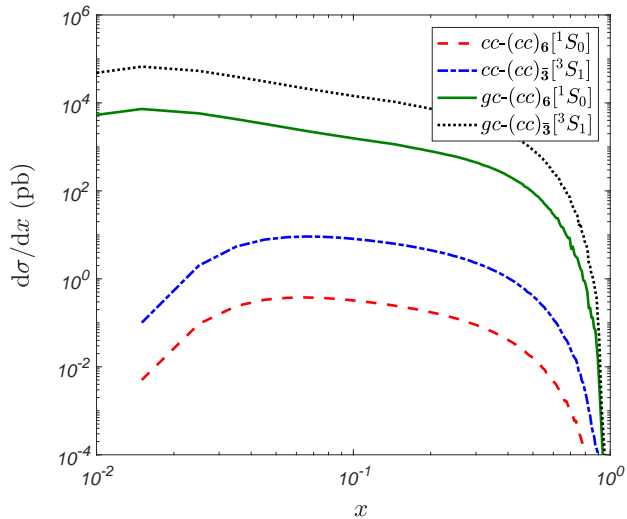}
\caption{The $x$ distributions of $\Xi_{cc}$ for various intermediate diquark states at the After@LHC with intrinsic charm component as $A_{\rm in}=1\%$. $p_t>0.2\;\rm GeV$. }  \label{x}
\end{figure}

We present the $\Xi_{cc}$ distributions at the After@LHC versus the transverse momentum ($p_t$), rapidity ($y$), and pseudo-rapidity ($\eta$) in Figs.~\ref{pt}, \ref{rap}, and \ref{psrap}, respectively. Those distributions are consistent with the results in Tables \ref{eppt} and \ref{zey}. To compare with Fig.~\ref{cptgg}, Fig.~\ref{pt} shows the $\Xi_{cc}$ production in small $p_t$ region is dominated by the $(g+c)$ channel, and the $(g+g)$ channel still dominates over the $(c+c)$ channel in almost the whole $p_t$ region. Figs.~\ref{rap} and \ref{psrap} show the plateaus of $|y|\le1.5$ and $|\eta|\le2$ appear in $c+c$ channel, which become broader in $g+c$ channel as $|y|\le3$ and $|\eta|\le3$. We plot $x$ distribution for the $\Xi_{cc}$ production in the $(g+c)$ and $(c+c)$ subprocesses as shown in Fig.~\ref{x}. Contributions from small $x$ range play the dominant role in the $\Xi_{cc}$ production both in $(g+c)$ and $(c+c)$ channels.

\begin{figure}[htb]
\includegraphics[width=0.5\textwidth]{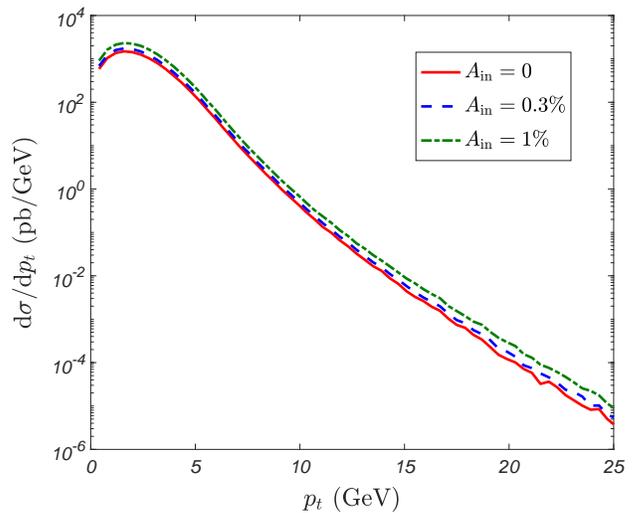}
\caption{The comparison of $p_t$ distributions for the hadroproduction of $\Xi_{cc}$ under different choices of $A_{\rm in}$ at the After@LHC, where contributions from various production schemes, i.e., $(g+g)$, $(g+c)$, and $(c+c)$, have been summed up. $p_t>0.2\;\rm GeV$ and no $y$ cut has been applied.}
\label{cpt}
\end{figure}

\begin{figure}[htb]
\includegraphics[width=0.5\textwidth]{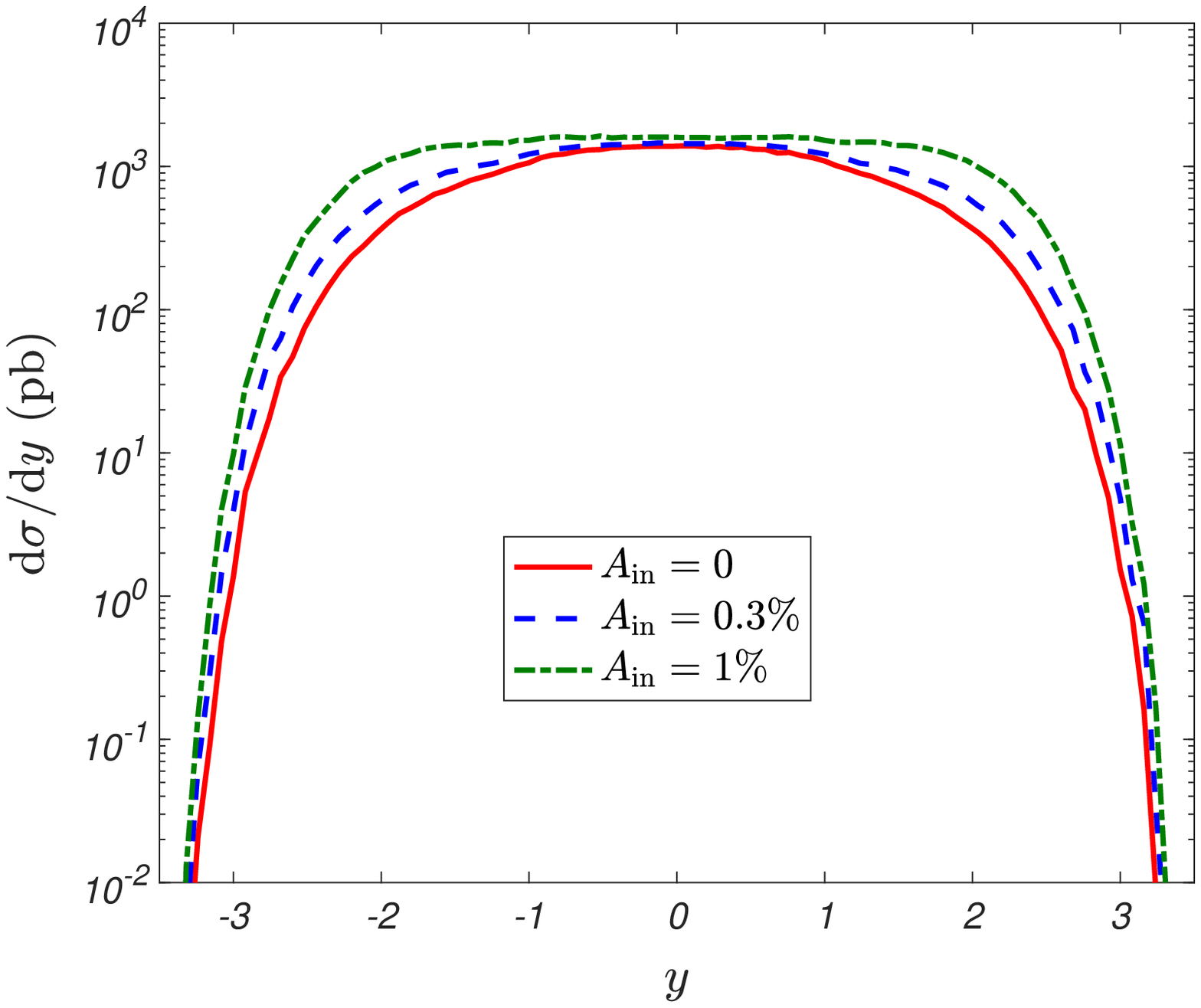}
\caption{The comparison of $y$ distributions for the hadroproduction of $\Xi_{cc}$ under different choices of $A_{\rm in}$ at the After@LHC, where contributions from various production schemes, i.e., $(g+g)$, $(g+c)$, and $(c+c)$, have been summed up. $p_t>0.2\;\rm GeV$ and no $y$ cut has been applied.}
\label{crap}
\end{figure}

\begin{figure}[htb]
\includegraphics[width=0.5\textwidth]{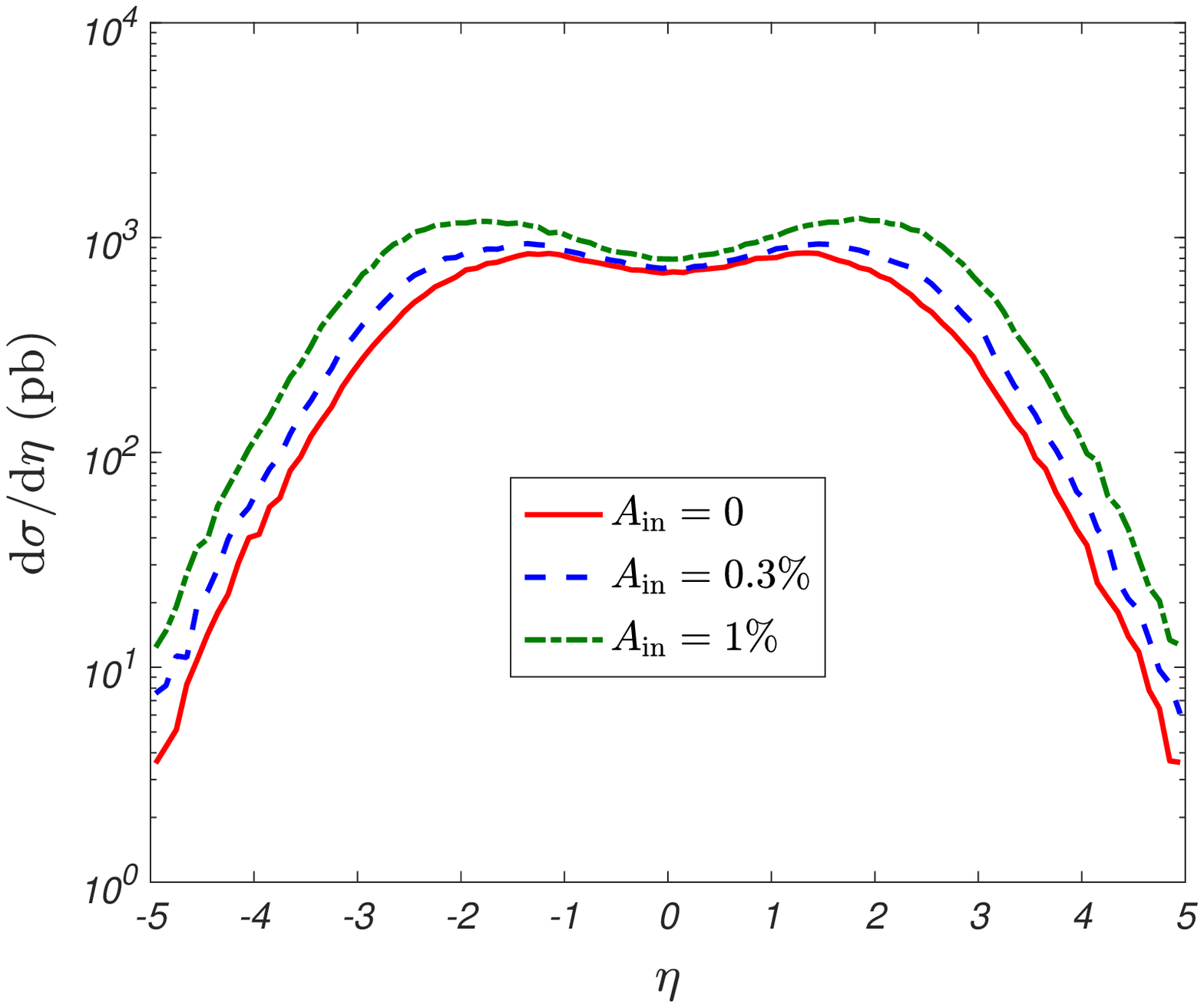}
\caption{The comparison of $\eta$ distributions for the hadroproduction of $\Xi_{cc}$ under different choices of $A_{\rm in}$ at the After@LHC, where contributions from various production schemes, i.e., $(g+g)$, $(g+c)$, and $(c+c)$, have been summed up. $p_t>0.2\;\rm GeV$ and no $y$ cut has been applied.}
\label{cpsrap}
\end{figure}

\begin{figure}[htb]
\includegraphics[width=0.5\textwidth]{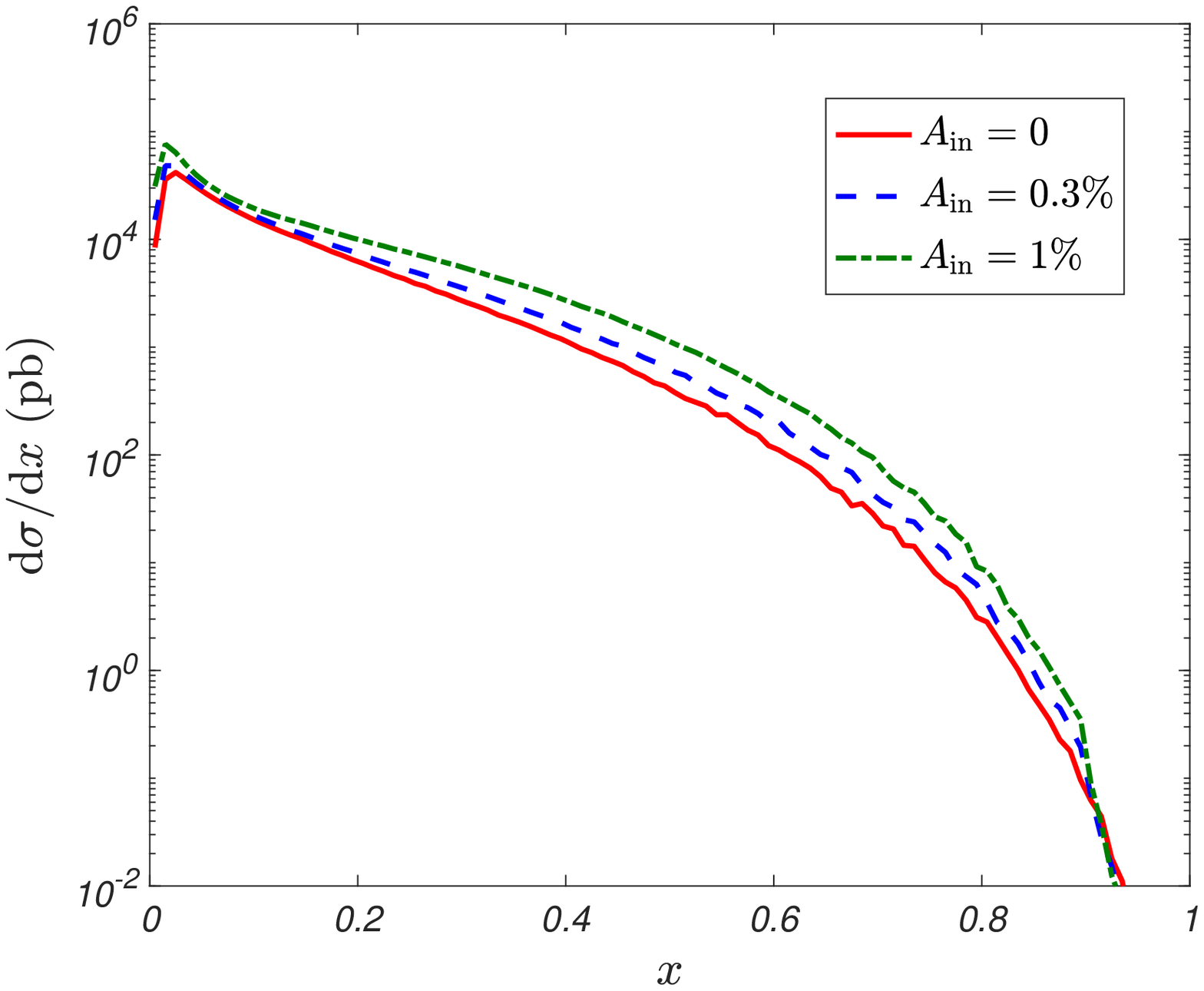}
\caption{The comparison of $x$ distributions for the hadroproduction of $\Xi_{cc}$ under different choices of $A_{\rm in}$ at the After@LHC, where contributions from various production schemes, i.e., $(g+g)$, $(g+c)$, and $(c+c)$, have been summed up. $p_t>0.2\;\rm GeV$ and no $y$ cut has been applied.}
\label{cx}
\end{figure}

\begin{figure}[htb]
\includegraphics[width=0.54\textwidth]{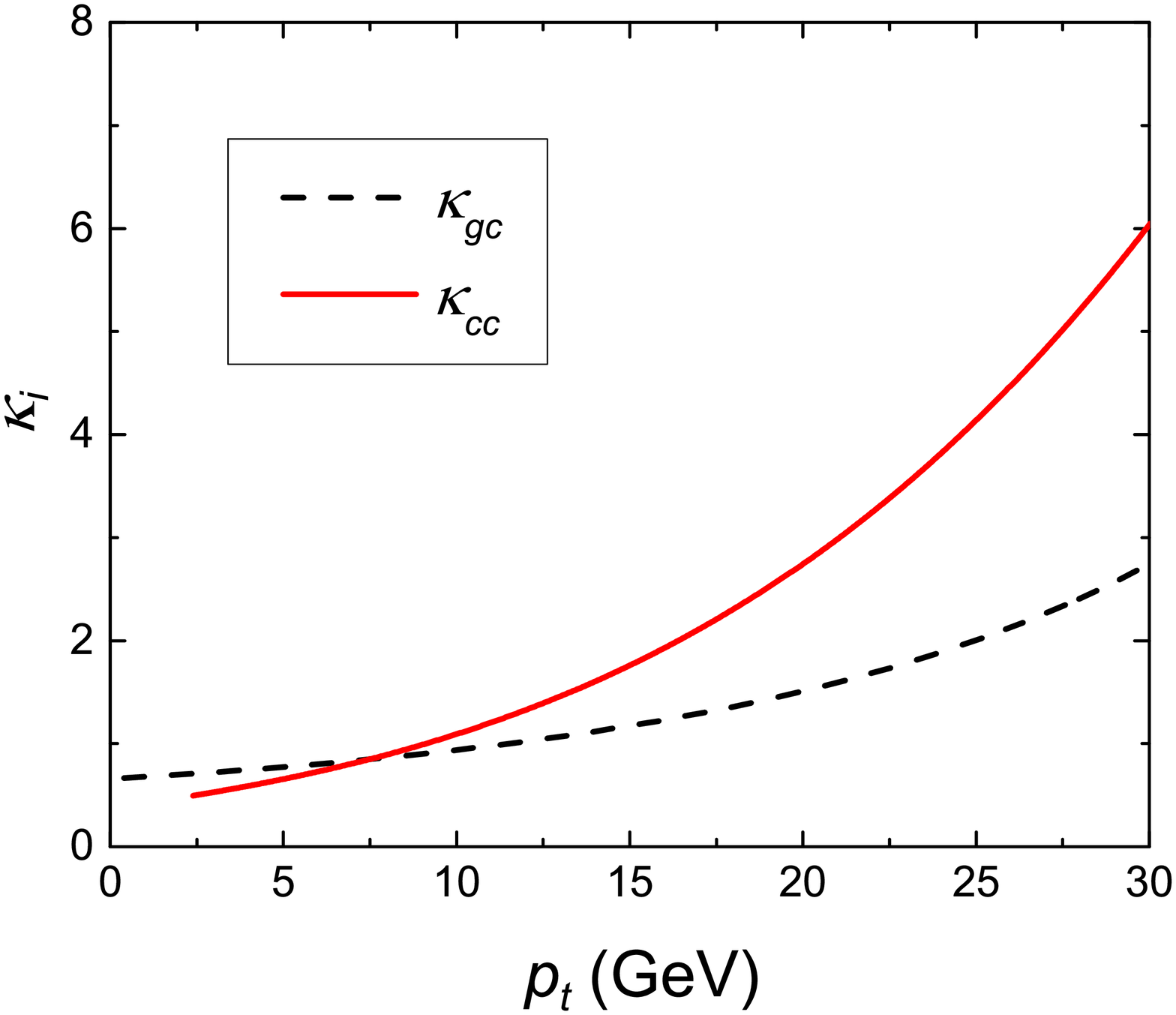}
\caption{The $\kappa_i$ ($i=g+c,c+c$) defined in Eq.~(\ref{kapp}) versus $p_t$ of $\Xi_{cc}$ with intrinsic charm component $A_{\rm in}=1\%$ at the After@LHC, in which contributions from different intermediate diquark states have been summed up. $p_t>0.2\;\rm GeV$ and no $y$ cut are applied.}
\label{kfig}
\end{figure}

\begin{figure}[htb]
\includegraphics[width=0.55\textwidth]{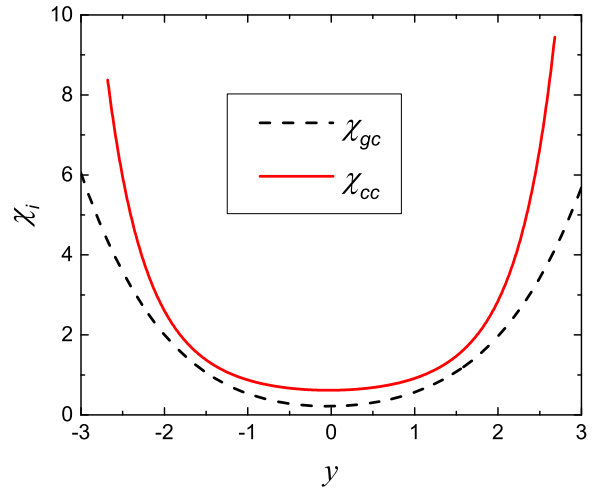}
\caption{The $\chi_i$ ($i=g+c,c+c$) defined in Eq.~(\ref{chi}) versus $y$ of $\Xi_{cc}$ with intrinsic charm component $A_{\rm in}=1\%$ at the After@LHC, in which contributions from different intermediate diquark states have been summed up. $p_t>0.2\;\rm GeV$ and no $y$ cut are applied.}
\label{chfig}
\end{figure}

To show how the intrinsic charm affects the differential distributions, we present the $p_t$, $y$, $\eta$, and $x$ distributions for $A_{\rm in}=0, 0.3\%, 1\%$ in Figs.~\ref{cpt},~\ref{crap}, and~\ref{cpsrap}, respectively. Here the contributions of $(cc)_{\bar{\bf 3}}[^3S_1]$ and $(cc)_{\bf 6}[^1S_0]$ configurations, and results from different production schemes, i.e., $(g+g)$, $(g+c)$, and $(c+c)$, have been summed up. The $p_t$ distributions are close in shape for various $A_{\rm in}$, however their differences become obvious in large $p_t$ region. The $y$ and $\eta$ distributions change more significantly with variation of $A_{\rm in}$ from $0$ to $1\%$. For example, both the shape and the normalization of $y$-distribution are changed significantly with the increment of $A_{\rm in}$. In Fig.~\ref{cx}, we present the comparison of $x$ distributions with different intrinsic charm component. It shows that the intrinsic charm provides contribution in large $x$ region, which is consistent with previous results as shown in Fig.~\ref{pdfc}. These changes of distributions are large enough to be potentially observed by the After@LHC for searching the intrinsic charm component in a proton.To show how the distributions change with the transverse momentum and rapidity, similar to the ratios $\varepsilon_i\,(p_{t\rm cut})$ and $\zeta_i\,(y_{\rm cut})$, we introduce two ratios $\kappa_i$ and $\chi_i$, i.e.ïŒ?\begin{equation}\label{kapp}
\kappa_i=\frac{d \sigma_{i}/dp_t-d \sigma^0_{i}/dp_t}{d \sigma^0_{i}/dp_t},
\end{equation}
and
\begin{equation}\label{chi}
\chi_i=\frac{d \sigma_{i}/dy-d \sigma^0_{i}/dy}{d \sigma^0_{i}/dy}.
\end{equation}
Here subscript $i$ stands for $g+c$ or $c+c$ mechanism, respectively. $\sigma$ denotes the cross section of $A_{\rm in}=1\%$ and $\sigma^0$ denotes that of $A_{\rm in}=0$, in which contributions of different diquark configuration have been summed up. The results are put in Figs.~\ref{kfig} and \ref{chfig}, which show in larger $p_t$ and larger rapidity regions, contribution from intrinsic charm are more obvious.

\subsection{Theoretical uncertainties for $\Xi_{cc}$ production}

In this subsection, we discuss the main theoretical uncertainties for the $\Xi_{cc}$ production at the After@LHC, which are from the choices of the charm quark mass, the renormalization scale, and the intrinsic charm PDF, respectively. When discussing the uncertainty from one error source, other input parameters shall be kept to be their central values. For convenience, we set $A_{\rm{in}}=1\%$ throughout this subsection.

\begin{table}[htb]
\begin{tabular}{|c|c|c|c|}
\hline
~$m_c$ (GeV) & ~1.65~ & ~1.75~ & ~1.85~ \\
\hline
 $g+g \to (cc)_{\bar{{\bf 3}}}[^3S_1]$ & $1.27\times10^3$ & $7.55\times 10^{2}$ & $4.57\times10^{2}$  \\
\hline
$g+g \to (cc)_{{\bf 6}}[^1S_0]$ & $2.32\times10^{2}$ & $1.37\times 10^{2}$ & $8.24\times10^{1}$ \\
\hline
 $g+c \to (cc)_{\bar{{\bf 3}}}[^3S_1]$ & $7.58\times10^3$ & $5.32\times 10^{3}$ & $3.76\times10^{3}$  \\
\hline
$g+c \to (cc)_{{\bf 6}}[^1S_0]$ & $8.22\times10^{2}$ & $5.78\times 10^{2}$ & $4.09\times10^{2}$ \\
\hline
 $c+c \to (cc)_{\bar{{\bf 3}}}[^3S_1]$ & 3.24 & 1.79 & 1.25  \\
\hline
$c+c \to (cc)_{{\bf 6}}[^1S_0]$ & $1.33\times10^{-1}$ & $7.16\times 10^{-2}$ & $5.12\times10^{-2}$ \\
\hline
\end{tabular}
\caption{Total cross sections (in unit pb) for the $\Xi_{cc}$ production at the After@LHC under different choices of $m_c$ mass. $p_t>0.2$ GeV and $A_{\rm{in}}=1\%$.}
\label{mass}
\end{table}

Total cross sections for $m_c=1.75\pm0.10\,\rm GeV$ are presented in Table~\ref{mass}, which shows
\begin{eqnarray}
\sigma_{g+g \to (cc)_{\bar{\bf 3}}[^3S_1]} &=& \left(7.55^{+5.15}_{-2.98}\right)\times10^{2}\;{\rm pb}, \nonumber\\
\sigma_{g+g \to (cc)_{\bf 6}[^1S_0]} &=& \left(1.37^{+0.95}_{-0.55}\right)\times10^{2}\;{\rm pb}, \nonumber\\
\sigma_{g+c \to (cc)_{\bar{\bf 3}}[^3S_1]} &=& \left(5.69^{+2.44}_{-1.68}\right)\times10^{3}\;{\rm pb}, \nonumber\\
\sigma_{g+c \to (cc)_{\bf 6}[^1S_0]} &=& \left(6.19^{+2.64}_{-1.82}\right)\times10^{2}\;{\rm pb}, \nonumber\\
\sigma_{c+c\to (cc)_{\bar{\bf 3}}[^3S_1]} &=& 2.02^{+1.61}_{-0.59}\;{\rm pb}, \nonumber\\
\sigma_{c+c\to (cc)_{\bf 6}[^1S_0]} &=& \left(8.03^{+6.77}_{-2.25}\right)\times10^{-2}\;{\rm pb}.
\end{eqnarray}
Total cross section depends heavily on the choice of charm quark mass, which shall be changed by $[-39\%,69\%]$ for $g+g$ channel, $[-30\%,43\%]$ for $g+c$ channel, and $[-29\%,84\%]$ for the $c+c$ channel, respectively.

\begin{table}[htb]
\begin{tabular}{|c|c|c|c|}
\hline
$\mu_{R}$ & $\sqrt{\hat{s}}$ & $\sqrt{\hat{s}}/2$ & $M_t$ \\
\hline
 $g+g \to (cc)_{\bar{{\bf 3}}}[^3S_1]$ & $1.63\times10^2$ & $3.99\times 10^{2}$ & $7.55\times10^{2}$  \\
\hline
$g+g \to (cc)_{{\bf 6}}[^1S_0]$ & $3.13\times10^{1}$ & $7.67\times 10^{1}$ & $1.37\times10^{2}$ \\
\hline
 $g+c \to (cc)_{\bar{{\bf 3}}}[^3S_1]$ & $3.43\times10^3$ & $5.47\times 10^{3}$ & $5.32\times10^{3}$  \\
\hline
$g+c \to (cc)_{{\bf 6}}[^1S_0]$ & $3.76\times10^{2}$ & $5.99\times 10^{2}$ & $5.78\times10^{2}$ \\
\hline
 $c+c \to (cc)_{\bar{{\bf 3}}}[^3S_1]$ & 1.25 & 1.76 & 1.79  \\
\hline
$c+c \to (cc)_{{\bf 6}}[^1S_0]$ & $5.05\times10^{-2}$ & $7.03\times 10^{-2}$ & $7.16\times10^{-2}$ \\
\hline
\end{tabular}
\caption{Total cross sections (in unit pb) for the $\Xi_{cc}$ production at the After@LHC under different choices of renormalization scale $\mu_R$. $p_t>0.2$ GeV and $A_{\rm{in}}=1\%$.}
\label{scale}
\end{table}

In the above estimations, we have fixed the renormalization scale $\mu_R$ to be the transverse mass of $\Xi_{cc}$, e.g., $m_{T}=\sqrt{p_t^2+M_{\Xi_{cc}}^2}$, which is usually adopted in the literature. Taking another two choices, e.g., $\mu_R=\sqrt{\hat{s}}/2$ and $\mu_R=\sqrt{\hat{s}}$, we estimate the renormalization scale uncertainty, where $\sqrt{\hat{s}}$ is the center-of-mass energy of the subprocess. Numerical results are presented in Table~\ref{scale}. For the case of $\Xi_{cc}$ production via $(g+c)$ channel, the scale uncertainty is about $\pm35\%$.

\begin{table}[htb]
\begin{center}
\begin{tabular}{|c|c|c|c|c|}
\hline
 & \multicolumn{2}{|c|} {$\sigma_{g+c}$ (pb)}& \multicolumn{2}{|c|} {$\sigma_{c+c}$ (pb)}\\
\hline
 & $(cc)_{\bar{{\bf 3}}}[^3S_1] $ & $(cc)_{{\bf 6}}[^1S_0] $  & $(cc)_{\bar{{\bf 3}}}[^3S_1] $ & $(cc)_{{\bf 6}}[^1S_0] $  \\
\hline
BHPS & $5.32\times 10^{3}$ & $5.78\times 10^{2}$ & 1.79 & $7.16\times 10^{-2}$  \\
\hline
CT14C-BHPS1 & $6.39\times 10^{3}$ & $6.95\times 10^{2}$ & 1.68 & $6.77\times 10^{-2}$  \\
\hline
CT14C-SEA1 & $6.79\times 10^{3}$ & $7.39\times 10^{2}$ & 1.26 & $5.15\times 10^{-2}$   \\
\hline
\end{tabular}
\caption{Total cross sections for three different intrinsic charm PDFs. CT14+BHPS is result by using the BHPS model evolved with Eq.(\ref{intrc}), CT14C-BHPS1 and CT14C-SEA1 are results for the CTEQ PDFs under BHPS model and SEA model~\cite{Hou:2017khm}, respectively. All the intrinsic charm PDFs are normalized to $1\%$. $p_t>0.2$ GeV.} \vskip 0.6cm
\label{updf}
\end{center}
\end{table}

To show how different models of IC PDF affect the production rates, we adopt the CTEQ PDF version CT14C under BHPS model, SEA model~\cite{Hou:2017khm} as explicit examples to estimate the errors caused by different choices of the IC PDF. The results are shown in Table~\ref{updf}. Both the CT14C-BHPS1 and CT14C-SEA1 are characterized by the magnitude of the intrinsic charm component by the first moment of the charm distribution $\langle x \rangle_{\rm IC}=0.57\%$, which corresponds to $1\%$ probability for finding intrinsic charm component in a proton. Table~\ref{updf} shows that by using those three IC PDFs, the total cross sections vary by about $20\% \sim 27\%$ and $6\% \sim 30\%$ for the $(g+c)$ and $(c+c)$ mechanisms, respectively.

\begin{table*}[htb]
\begin{center}
\begin{tabular}{|c|c|c|c|c|}
\hline
 & \multicolumn{2}{|c|} {$\sigma_{g+c}$ (pb)}& \multicolumn{2}{|c|} {$\sigma_{c+c}$ (pb)}\\
\hline
 & $(cc)_{\bar{{\bf 3}}}[^3S_1] $ & $(cc)_{{\bf 6}}[^1S_0] $  & $(cc)_{\bar{{\bf 3}}}[^3S_1] $ & $(cc)_{{\bf 6}}[^1S_0] $  \\
\hline
NNPDF3IC-1330 & $4.85\times 10^{3}$ & $5.27\times 10^{2}$ & 1.49 & $5.97\times 10^{-2}$   \\
\hline
NNPDF3IC-1610 & $4.49\times 10^{3}$ & $4.86\times 10^{2}$ & 1.45 & $5.78\times 10^{-2}$   \\
\hline
CT10C-BHPS1 & $5.99\times 10^{3}$ & $6.51\times 10^{2}$ & 1.50 & $6.03\times 10^{-2}$   \\
\hline
CT10C-SEA1 & $6.33\times 10^{3}$ & $6.88\times 10^{2}$ & 1.14 & $4.67\times 10^{-2}$   \\
\hline
\end{tabular}
\caption{Total cross sections for different choices of intrinsic charm PDF with various IC models. $p_t>0.2$ GeV.} \vskip 0.6cm
\label{updf2}
\end{center}
\end{table*}

As a final remark, if choosing the recently developed model independent NNPDF3IC~\cite{Ball:2016neh} as the input for the IC PDF, whose input parameters are based on a NLO calculation and are fixed via a global fitting of experimental data of deep inelastic structure functions, we shall obtain a slightly smaller total cross-sections than the cases of CT14+BHPS and CT14C-BHPS1~\footnote{A smaller total cross-section is reasonable, since the fitted NNPDF3IC prefers a scale-dependent probability of finding IC component in a proton, e.g. $0.7\%\pm0.3\%$ for the scale equals to $1.65$ GeV~\cite{Ball:2016neh}, which is smaller than our present choice of $1\%$ for CT14+BHPS and CT14C-BHPS1; More over, the NNPDF3IC PDF becomes negative for $x \gtrsim 0.75$.}. The NNPDF3IC results are presented in Tab.~\ref{updf2}, which are for the NNPDF3IC preferable $m_c$ range of $[1.33, 1.61]$ GeV.

\section{Conclusions}\label{summary}

In the paper, we have studied the hadronic production of $\Xi_{cc}$ baryon at the fixed-target experiment at the LHC, e.g. After@LHC. More accurate data are assumed to be available at the After@LHC than the SELEX experiment, which shall be helpful to clarify the previous SELEX puzzle on the $\Xi_{cc}$ production. Our results show that the intrinsic charm can have significant impact on the $\Xi_{cc}$ production. If setting the probability of finding the intrinsic charm in proton is $A_{\rm in}=1\%$, the total production cross section can be enhanced by a factor of $2$ through the $(g+c)$ and $(c+c)$ channels. By summing up contributions from $(g+g)$, $(g+c)$, and $(c+c)$ channels and contributions from both diquark states $(cc)_{\bar{\bf 3}}[^3S_1]$ and $(cc)_{\bf 6}[^1S_0]$, we shall have $3.4\times10^5$ or $1.4\times10^7$ $\Xi_{cc}$ events per operation year with the integrated luminosity $0.05\,{\rm fb}^{-1}$ or $2\,{\rm fb}^{-1}$, respectively.

Thus, the fixed-target experiment After@LHC can be an ideal platform for studying properties of $\Xi_{cc}$. Since the total cross sections and the differential distributions are sensitive to the probability of finding intrinsic charm component in a proton, the After@LHC shall also be a good platform for testing the intrinsic charm mechanism and for fixing the intrinsic charm PDF.

\hspace{2cm}

{\bf Acknowledgements}: We thank Hua-Yong Han and Yun-Qing Tang for helpful discussions on the intrinsic charm PDF. This work was supported in part by the Natural Science Foundation of China under Grant No.11605029, No.11625520, and No.11847301, and by the Fundamental Research Funds for the Central Universities under Grant No.2019CDJDWL0005.


\end{document}